\documentclass[prl,letterpaper,twocolumn]{revtex4}%
\usepackage{amsfonts}
\usepackage{amsmath}
\usepackage{amssymb}
\usepackage{graphicx}%
\providecommand{\U}[1]{\protect\rule{.1in}{.1in}}

\begin{document}
\title{Magnetic Field Dependence and Efimov Resonance Broadening in Ultracold
Three-Body Recombination}
\author{Seth T. Rittenhouse}
\affiliation{ITAMP, Harvard-Smithsonian Center for Astrophysics, Cambridge, MA 02138}
\email{\texttt{srittenhouse@cfa.harvard.edu}}

\begin{abstract}
We derive an analytic formula which describes the final bound state dependence
in ultracold three-body recombination. Using an energy-dependent loss
parameter, the recently observed broad resonance in an ultracold gas of $^{6}%
$Li atoms \cite{ottenstein2008cst,huckans2009tbr} is described quantitatively.
We also provide an analytic and approximation for the three-body recombination
rate which encapsulates the underlying physics of the universal three-body
recombination process.

\end{abstract}
\date{\today
}
\maketitle

Over the last several years, the field of Efimov physics has generated a great
deal of excitement \cite{efimov1970ela,efimov1973}. Ultracold atomic gases
provide a unique environment for studying these exotic states. Through the use
of a broad Fano-Feshbach resonance the s-wave scattering length can be tuned
over several orders of magnitude
\cite{cornish2000srb,bartenstein2004pdl2,pollack2009eti}. Using these tools
many universal behaviors associated with Efimov physics have been observed in
both three- and four-body recombination processes
\cite{knoop2009oel,barontini2009oha,zaccanti2009oes,ferlaino2009euf,pollack2009utf}%
, largely confirming many theoretical predictions (see Refs.
\cite{esry1999rta,bedaque2000tbr,vonstecher2009suf,mehta2009gtd} for some
examples). The majority of these experimental studies have taken place using
an ultracold gas of bosonic atoms.

Recently, a new class of Efimov state has emerged in a three-component
degenerate Fermi gas of $^{6}$Li atoms. For magnetic fields between 30 and 600
G, two resonances in three-body recombination have been observed
\cite{ottenstein2008cst,huckans2009tbr}. The positions of these resonances
corresponds to a single three-body bound state crossing the continuum at lower
and then again at higher field strength. While the position of these
resonances is fit very well by existing three-body recombination theory
\cite{braaten2009three,naidon2009possible,floerchinger2009frt}, the widths are
not. The standard theory, in which it is assumed that the short range behavior
of the three-body system is largely independent of the strength of the
external magnetic field, predicts that the two resonances should have roughly
the same width. The experimental evidence show a higher field resonance that
is significantly broader than the first.

Three-body recombination occurs when two particles combine to form a dimer
state releasing the resulting binding energy in the form of kinetic energy
between the dimer and a third particle. It has been proposed by Wenz
\textit{et al.} \cite{wenz2009utt} that the binding energy $E_{b}$ of the
final dimer state is to blame for the observed discrepancy in resonance
widths. The binding energy of the first several deeply bound two-body states
in $^{6}$Li have a strong magnetic field dependence \cite{wenz2009utt}. By
assuming a $1/E_{b}$ dependence in the loss parameter they were able to find
reasonable agreement with both resonances. The question still remains,
however, as to where this energy dependence comes from and whether 1/$E_{b}$
is the correct form.

\begin{figure}[th]
\begin{center}
\includegraphics[width=3in]{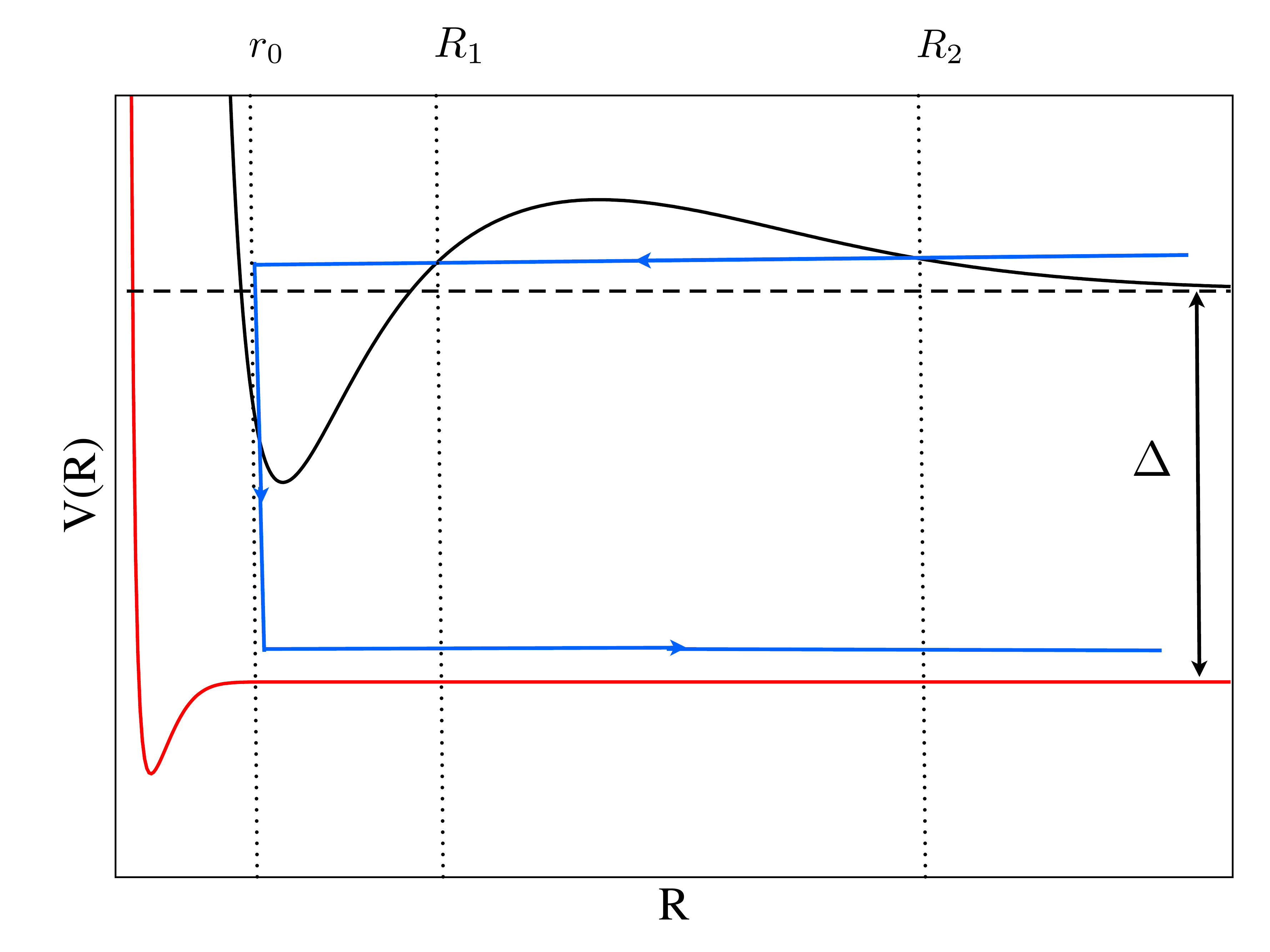}
\end{center}
\caption{(color online) A Schematic of the two-channel scattering process that
controls threshold three-body recombination is shown.}%
\label{Fig:Chann_Schem}%
\end{figure}

In this paper we present a simple mechanism that describes the broadening of a
three-body recombination resonance due to the magnetic field dependence of a
deeply bound two-body state. We proceed by assuming that this precess can be
described by a two-channel inelastic scattering process within the adiabatic
hyperspherical representation, shown schematically in Fig.
(\ref{Fig:Chann_Schem}). The upper channel consists of three free particles
scattering near threshold, while the lower channel consists of a tightly bound
dimer with binding energy $\Delta$ and a free particle. The two channels are
coupled at some small hyperradius $r_{0}$ by a non-adiabatic coupling
$P$-matrix element:%
\begin{equation}
P_{if}\left(  R\right)  =\left\langle \Phi_{f}\left(  R;\Omega\right)
\left\vert d/dR\right\vert \Phi_{i}\left(  R;\Omega\right)  \right\rangle
_{\Omega}, \label{Eq:P-mat}%
\end{equation}
where $\Phi_{i\left(  f\right)  }\left(  R;\Omega\right)  $ is the
hyperangular channel function describing the initial (final) adiabatic channel
and the matrix element is taken over the hyperangular degrees of freedom.

Mehta \textit{et al.} \cite{mehta2009gtd} showed that the inelastic scattering
rate at the three-body threshold for this type of system can be described
within the Wentzel--Kramers--Brillouin (WKB) approximation as%
\begin{equation}
K_{3}=\dfrac{9\hbar^{5}}{4m^{3}E^{2}}\dfrac{Ce^{-2\gamma_{WKB}}\sinh2\eta
}{\cos^{2}\phi_{WKB}+\sinh^{2}\eta}, \label{Eq:rec_rate1}%
\end{equation}
where $m$ is the atom mass, $E$ is the initial asymptotic kinetic energy of
the three atoms, and the constant $C$ is determined below. In Eq.
(\ref{Eq:rec_rate1}) $\gamma_{WKB}$ is the WKB tunneling integral and
$\phi_{WKB}$ is the WKB phase accumulated in the inner potential well of the
upper potential, i.e.%
\begin{align}
\gamma_{WKB}  &  =\operatorname{Re}\left[  \int_{r_{0}}^{R_{2}}\sqrt{\left[
\dfrac{2\mu}{\hbar^{2}}\left(  V_{i}\left(  R\right)  -E\right)  +\dfrac
{1}{4R^{2}}\right]  }dR\right]  ,\label{Eq::WKB_tunn}\\
\phi_{WKB}  &  =\operatorname{Im}\left[  \int_{r_{0}}^{R_{2}}\sqrt{\left[
\dfrac{2\mu}{\hbar^{2}}\left(  V_{i}\left(  R\right)  -E\right)  +\dfrac
{1}{4R^{2}}\right]  }dR\right]  . \label{Eq:WKB_phase}%
\end{align}
where $V_{i}\left(  R\right)  $ is the initial adiabatic potential which can
be found using known methods (see for instance Ref. \cite{nielsen2001tbp}) and
$R_{1}$ is the outer classical turning point, and $E$ is the initial
asymptotic kinetic energy of the three atoms. The extra $1/4R^{2}$ in Eqs.
(\ref{Eq::WKB_tunn}) and (\ref{Eq:WKB_phase}) is due to the Langer correction.
The proportionality constant $C\approx300$ in Eq. (\ref{Eq:rec_rate1}) is
extracted by comparing the $K_{3}$ in case when all three scattering lengths
are equal to $a$, where $\left\vert a\right\vert \gg r_{0}$, to the known
recombination rate for this case from Ref. \cite{Braaten2006physrep}:%
\begin{equation}
K_{3}=\dfrac{\hbar4590}{m}\dfrac{a^{4}\sinh2\eta}{\cos^{2}\phi_{WKB}+\sinh
^{2}\eta}.
\end{equation}

In Eq. (\ref{Eq:rec_rate1}), $\eta$ is an imaginary phase which parametrizes
the losses from the initial channel. The ratio of the outgoing to incoming
probabilities in the initial channel \cite{Braaten2006physrep} is given in
terms of $\eta$ as
\begin{equation}
\dfrac{\left\vert \Psi_{out}\right\vert ^{2}}{\left\vert \Psi_{in}\right\vert
^{2}}=e^{-4\eta}. \label{Eq:eta_def}%
\end{equation}
The constants $\eta\ $and $r_{0}$ are determined by short range physics and
are usually used to fit the position and width of an Efimov resonance
\cite{braaten2009three,naidon2009possible,floerchinger2009frt,dincao2009ultracold}%
. For three-body recombination of distinguishable $^{6}$Li atoms, assuming
that $\eta$ and $r_{0}$ are fixed over the experimental range of magnetic
fields, there is excellent agreement with the first three-body resonance
\cite{braaten2009three,naidon2009possible,floerchinger2009frt} at
approximately 130 G. A problem occurs however at the second resonance; the
theory prediction is far narrower than the experimental results.

By assuming that the $P$-matrix element coupling the two channels is
Lorentzian which peaks at $R=r_{0}$, the energy dependence of $\eta$ can be
extracted. By considering the Landau-Zener transition probability, Clark
\cite{clark1979cna} has shown that in this case the probability of making a
non-adiabatic transition as the system passes through the transition region is
given by%
\begin{align}
P_{na}  &  =e^{-2\pi\gamma},\label{Eq:probability}\\
\gamma &  =\dfrac{1}{\hbar v}\dfrac{\Delta}{8P_{\max}},\nonumber
\end{align}
where $v$ is the characteristic velocity of the system at $R=r_{0}$, $\Delta$
is the energy separating the initial and final channels and $P_{\max}$ is the
maximum of the $P$-matrix element. The velocity $v$ is determined by the
short-range physics of the incoming channel while $P_{\max}$ will be
determined by the small $R$ behavior of both the incoming and outgoing channel
function $\Phi_{i}$ and $\Phi_{f}.$

While $P_{\max}$ cannot be determined exactly, we can make certain statements
about its behavior. We assume that the short-range nature of the incoming
channel is independent of the magnetic field over the range we consider. The
outgoing final channel will be shifted overall, but the size and wave function
of the final deeply-bound dimer state is relatively unaffected. Thus, both
$P_{\max}$ and $v$ are independent of the final state energy $\Delta.$

By assuming that any non-adiabatic transition results in a three-body
recombination event and the system must pass twice through the transition, on
the way in and on the way out, the probability of remaining in the initial
channel can be extracted:%
\begin{equation}
\dfrac{\left\vert \Psi_{out}\right\vert ^{2}}{\left\vert \Psi_{in}\right\vert
^{2}}=\left(  1-P_{na}\right)  ^{2}. \label{Eq:Prob_remains}%
\end{equation}
Comparing Eqs. (\ref{Eq:eta_def}) and (\ref{Eq:Prob_remains}) gives a simple
equation that can be solved for $\eta$:%
\begin{align}
\eta &  =\dfrac{1}{2}\ln\left[  \dfrac{1}{1-\exp\left(  -\beta\Delta\right)
}\right]  ,\label{Eq:eta_dep}\\
\beta &  =\dfrac{\pi}{\hbar v}\dfrac{1}{4P_{\max}}.\nonumber
\end{align}
With this formula, the unknown short-range dependence of $\eta$ is encompassed
in a single parameter $\beta$ which is independent of the binding energy
$\Delta.$ It is important to note that we have not increased the number of
fitting parameters, we have merely shifted the fitting to the system dependent
parameter $\beta$ instead of $\eta$. In cases where the final state energy
does not strongly depend on the magnetic field, this extra dependence is not
needed and $\eta$ can be used directly. However in cases where the final state
has a strong field dependence, such as $^{6}$Li, the proposed parametrization
of $\eta$ is more appropriate. In general a smaller binding energy leads to a
larger loss parameter in Eq. (\ref{Eq:eta_dep}), and a larger loss parameter
leads to a broader resonance. Qualitatively, while the binding energy
dependence in Eq. (\ref{Eq:eta_def}) is similar to the 1/$\Delta$ dependence
assumed in Ref. \cite{wenz2009utt}, the detailed behavior is considerably
different.\begin{figure}[t]
\begin{center}
\includegraphics[width=3in]{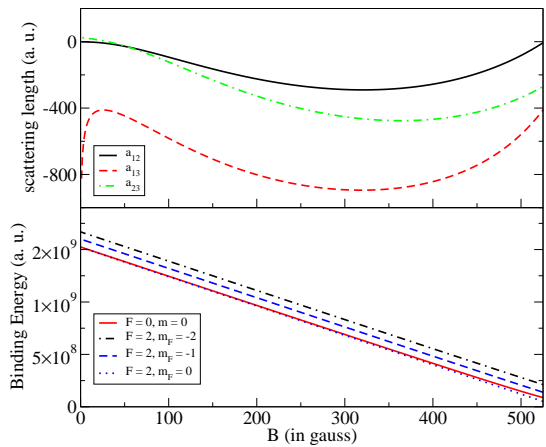}
\end{center}
\caption{(color online) (a) The three s-wave scattering lengths for $^{6}$Li
are shown in atomic units as a function of magnetic field. (b) The four
smallest dimer binding energies from Ref. \cite{wenz2009utt} are shown in
atomic units as a function of magnetic field. The total nuclear angular
momentum and projection are given, but are unimportant for the results in this
work.}%
\label{Fig:Scat_props}%
\end{figure}

\begin{figure}[t]
\begin{center}
\includegraphics[width=3in]{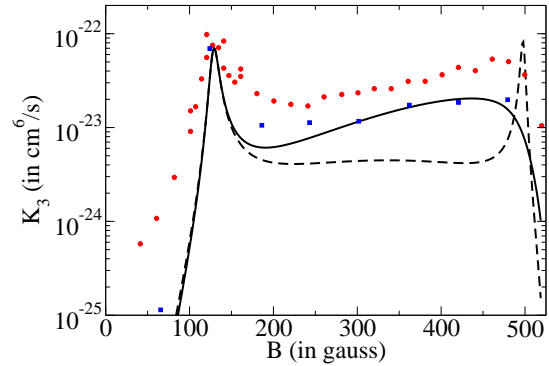}
\end{center}
\caption{(color online) The recombination rate constant predicted from Eq.
(\ref{Eq:rec_rate1}) is shown for fixed $\eta$ (dashed curve) and for energy
dependent $\eta$ (solid curve) as a function of magnetic field compared to
experimental data from Ref. \cite{ottenstein2008cst} (red circles) and Ref.
\cite{huckans2009tbr} (blue squares). }%
\label{Fig:Rec_Rate_comp}%
\end{figure}

Figure (\ref{Fig:Rec_Rate_comp}) shows the recombination rates predicted from
Eq. (\ref{Eq:rec_rate1}) in the threshold regime (incident energy $E=10^{-12}$
a. u.) compared to the experimental results of Ottenstein \textit{et al.}
\cite{ottenstein2008cst} (red circles) and Huckans \textit{et al.}
\cite{huckans2009tbr} (blue squares). The dashed curve is found by assuming
that $\eta$ is completely independent of the magnetic field strength. The
solid curve is found by using Eq. (\ref{Eq:eta_dep}) with the lowest dimer
binding energy from Fig. (\ref{Fig:Scat_props}b). Any one of these four
binding energies could have been used (with appropriate modifications to
$\beta$), but because they all have similar magnetic field dependence, the
results are nearly identical. The fits were found by setting $r_{0}=22$ bohr
such that the the first resonance occurs at 130 G. The dashed curve was found
by setting $\eta=0.05$ to fit the width of the first resonance at $B=130$ G
using the data from Ref. \cite{huckans2009tbr}. The solid curve was found by
choosing $\beta$ so that $\eta=exp\left(  -\beta\Delta\right)  =0.05$ at the
same $B$. Both of the predictions in Fig. (\ref{Fig:Rec_Rate_comp}) do an
excellent job of describing the first resonance, while using the $\Delta$
dependent loss parameter from Eq. (\ref{Eq:eta_dep}) is in astonishingly good
agreement with Huckans \textit{et al. }\cite{huckans2009tbr} and fairly good
agreement with Ottenstein \textit{et al.} \cite{ottenstein2008cst}.

The initial hyperradial potential used here was derived assuming that all
scattering lengths were much larger than any short range parameters. The
hyperradial potential used in Eqs. (\ref{Eq::WKB_tunn}) and
(\ref{Eq:WKB_phase}) were found by assuming zero range interactions, which is
appropriate for scattering lengths much greater than the size of the two-bod
interaction. In the case of $^{6}$Li, this size is given by approximately the
Van der Waals length $r_{6}\approx30$ bohr. Because the smallest scattering
length is not too much larger than this, we might expect small corrections to
this due to non-universal behavior in the potentials. Even with this caveat,
the qualitative agreement seen in Fig. (\ref{Fig:Rec_Rate_comp}) is
remarkable.

While the treatment above does give good agreement with experimental data, it
not necessarily convenient for quick comparison to experiment. Examining the
three scattering lengths from Fig. (\ref{Fig:Scat_props}a) shows that they
differ by a factor of roughly 2 throughout the region in which the
experimental data is taken. With this, we assume that they are different
enough to use the results of Ref. \cite{dincao2009ultracold} which gives the
incoming three-body hyperradial potential for three distinguishable particles
with large s-wave scattering lengths in four regions:%
\begin{equation}
V_{i}\left(  R\right)  =\left\{
\begin{array}
[c]{cc}%
\dfrac{\hbar^{2}}{2\mu}\dfrac{4-1/4}{R^{2}}, & \left\vert a_{l}\right\vert \ll
R\\
\dfrac{\hbar^{2}}{2\mu}\dfrac{4-1/4}{R^{2}}, & \left\vert a_{m}\right\vert \ll
R\ll\left\vert a_{l}\right\vert \\
\dfrac{\hbar^{2}}{2\mu}\dfrac{-s_{1}^{2}-1/4}{R^{2}}, & \left\vert
a_{s}\right\vert \ll R\ll\left\vert a_{m}\right\vert \\
\dfrac{\hbar^{2}}{2\mu}\dfrac{-s_{0}^{2}-1/4}{R^{2}}, & r_{0}\ll
R\ll\left\vert a_{s}\right\vert
\end{array}
\right.  .
\end{equation}
Here $a_{l},$ $a_{m}$ and $a_{s}$ are respectively the largest, second largest
and smallest scattering lengths in magnitude, and $s_{0}=1.006$ and
$s_{1}=0.414$ are parameters which are determined by the universal potential
in the limits where $R\ll\left\vert a_{s}\right\vert ,\left\vert
a_{m}\right\vert ,\left\vert a_{l}\right\vert $ and $\left\vert a_{s}%
\right\vert \ll R\ll\left\vert a_{m}\right\vert ,\left\vert a_{l}\right\vert $
respectively. It is not clear that these regimes exist for the scattering
lengths shown in Fig. (\ref{Fig:Scat_props}a), but we will proceed assuming
that they do. By assuming that transitions between different universal
behaviors have no significant contribution, the tunneling suppression and
phase accumulation can be found using Eqs. (\ref{Eq::WKB_tunn}) and
(\ref{Eq:WKB_phase}):%
\begin{align}
e^{-2\gamma_{WKB}}  &  \propto\left(  a_{l}a_{m}\right)  ^{2}%
,\label{Eq:univ_tunn}\\
\phi_{WKB}  &  =s_{0}\ln\left(  \left\vert a_{m}\right\vert /\left\vert
a_{s}\right\vert \right)  +s_{1}\ln\left(  \left\vert a_{s}\right\vert
/r_{0}\right)  ,\label{Eq:univ_phase}\\
s_{0}  &  =1.006;\text{ }s_{1}=0.414,\nonumber
\end{align}
Inserting this into Eq. (\ref{Eq:rec_rate1}) yields an analytic formula for
the three-body recombination rate:%
\begin{equation}
K_{3}^{univ}=\dfrac{\hbar}{m}\dfrac{C^{univ}\left(  a_{l}a_{m}\right)
^{2}\sinh2\eta}{\cos^{2}\left(  s_{0}\ln\dfrac{\left\vert a_{m}\right\vert
}{\left\vert a_{s}\right\vert }+s_{1}\ln\dfrac{\left\vert a_{s}\right\vert
}{r_{0}}\right)  +\sinh^{2}\eta}. \label{Eq:rec_rate2}%
\end{equation}
\begin{figure}[tb]
\begin{center}
\includegraphics[width=3in]{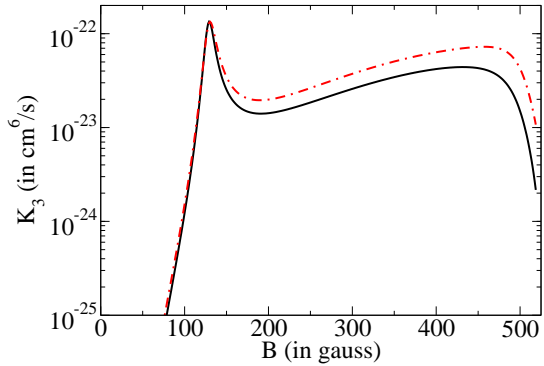}
\end{center}
\caption{The numerically calculated recombination rate constant from Fig.
(\ref{Fig:Rec_Rate_comp}) (solid curve) compared to the analytic approximation
from Eq. (\ref{Eq:rec_rate2}) is shown as a function of magnetic field
strength.}%
\label{Fig:Univ_vs_Num}%
\end{figure}

Figure (\ref{Fig:Univ_vs_Num}) shows the results of this formula compared to
the numerically obtained results in from Fig. (\ref{Fig:Rec_Rate_comp}). Both
rates use the same $\Delta$-dependent loss parameter, while we have set
$r_{0}=32.6$ bohr is Eq. (\ref{Eq:rec_rate2}) to fit the resonance position at
$B=130$ G. The constant $C^{univ}=1379$ has been chosen so that Eq.
(\ref{Eq:rec_rate2}) agrees with the numerical results at the peak of the
resonance. While the agreement may not be perfect it indicating that the
analytic expression from Eq. (\ref{Eq:rec_rate2}) encompasses the majority of
the relevant physics.

In summary, we have derived a simple expression that gives the loss parameter
$\eta$ as a function of the binding energy of the final outgoing states. While
this work focused on the three-body recombination resonances found in a
three-component degenerate Fermi gas of $^{6}$Li atoms, the method can be
applied to other three-body systems. For instance in the case of three-body
recombination to weekly bound dimers, the $P$-matrix and characteristic
velocity $v$ are known, and Eq. (\ref{Eq:eta_dep}) could be used to give a
more complete description of the process. This parametrization of $\eta$
introduces no extra fitting parameters and gives excellent agreement with
experiment. We have shown how a universal Efimov resonance can be broadened by
the a field dependent binding energy of the outgoing final state. In general,
a smaller dimer binding energy produces broader resonances when other
parameters are held fixed. We have also given an analytic expression for the
three-body recombination rate in a three-component degenerate Fermi gas of
$^{6}$Li atoms which provides reasonable agreement with the more complex
numerical results. The energy dependent loss parameter was derived here using
the intuition gained from the adiabatic hyperspherical method, but it is not
limited to this approach and could be used wherever N-body recombination
processes are parametrized by an imaginary phase. The process of determining
the width of loss resonances based on knowledge of the dimer binding energy
might also be turned around. One could use experimentally determined resonance
behavior to find the field dependence of deeply bound final states.

The author would like to thank S. Jochim and H. R. Sadeghpour for useful
discussions. Funding for this work was provided by a NSF through ITAMP at
Harvard University and Smithsonian Astrophysical Observatory.


\end{document}